\documentclass[12pt]{article}

\newcommand{\rmd}{{\rm d}}
\newcommand{\rmi}{{\rm i}}
\newcommand{\rme}{{\rm e}}

\usepackage{graphicx,cite}
\usepackage{epsf,latexsym}
\usepackage{amssymb,amsmath}

\begin{document}

\def\llra{\relbar\joinrel\longrightarrow}              
\def\mapright#1{\smash{\mathop{\llra}\limits_{#1}}}    
\def\mapup#1{\smash{\mathop{\llra}\limits^{#1}}}     
\def\mapupdown#1#2{\smash{\mathop{\llra}\limits^{#1}_{#2}}} 

\title{\bf The resonance amplitude associated with the Gamow states}

\author{Rafael de la Madrid \\
\small{\it Department of Physics, The Ohio State University at Newark,
Newark, OH 43055} \\
\small{E-mail: \texttt{rafa@mps.ohio-state.edu}}}

\date{}

\maketitle

\begin{abstract}
\noindent The Gamow states describe the quasinormal modes of 
quantum systems. It is shown that the resonance amplitude associated with the
Gamow states is given by the complex delta function. It is also shown that 
under the near-resonance approximation of neglecting the lower bound of the 
energy, such resonance amplitude becomes the Breit-Wigner amplitude. This
result establishes the precise connection between the Gamow states,
Nakanishi's complex delta function and the Breit-Wigner amplitude. In addition,
this result provides another theoretical basis for the 
phenomenological fact that the almost-Lorentzian peaks in cross sections 
are produced by intermediate, unstable particles.
\end{abstract}



\section{Introduction}
\setcounter{equation}{0}
\label{sec:Intro}

Resonances appear in all areas of quantum physics, in both the relativistic
and non-relativistic regimes. Resonances are intrinsic properties of a 
quantum system, and they describe the system's preferred ways of 
decaying. Experimentally, resonances appear as sharp peaks 
in the cross section that resemble the Breit-Wigner (Lorentzian) lineshape.

The Gamow states are the natural wave functions of resonances, and they were 
introduced by Gamow in his paper on $\alpha$-decay of radioactive 
nuclei~\cite{GAMOW}. Since then, they have been used by a number of authors, 
see e.g.~\cite{SIEGERT,PEIERLS,HUMBLET,ZELDOVICH,BERGGREN,GASTON,BERGGREN78,SUDARSHAN,
MONDRAGON83,CURUTCHET,BL,LIND,BERGGREN96,BOLLINI,FERREIRA,BETAN,MICHEL1,AJP02,KAPUSCIK1,MONDRAGON03,
MICHEL2,KAPUSCIK2,MICHEL3,MICHEL4,MICHEL5,URRIES,MICHEL6,TOMIO}. Likewise 
the bound 
states, the Gamow states are properties of the Hamiltonian, and they are 
associated with the natural frequencies of the system. The usefulness
of the Gamow states is attested by the remarkable success of the Gamow Shell 
Model~\cite{MICHEL1,MICHEL2,MICHEL3,MICHEL4,MICHEL5,MICHEL6} and similar
nuclear-structure formalisms~\cite{FERREIRA,BETAN}.

Because resonances leave a quasi-Lorentzian fingerprint in the cross section,
and because the Gamow states are the natural wave functions of resonances,
the resonance amplitude associated with the Gamow states must be related
to the Breit-Wigner amplitude. The purpose of this paper is to show that 
the resonance amplitude associated with the Gamow states 
is proportional to the complex delta function,
$\delta (E-z_{\rm R})$, and that such amplitude can be
approximated in the near-resonance region by the Breit-Wigner amplitude. More
precisely, we will show that the transition amplitude from a resonance 
state of energy $z_{\rm R}$ to a scattering state of energy $E\geq 0$, 
${\cal A}(z_{\rm R} \to E)$, is given by
\begin{equation}
  {\cal A}(z_{\rm R} \to E)= \rmi \sqrt{2\pi}{\cal N}_{\rm R} 
             \delta (E-z_{\rm R}) 
     \simeq -\frac{{\cal N}_{\rm R}}{\sqrt{2\pi}} \frac{1}{E-z_{\rm R}}  \, ,
    \quad E\geq 0 \, ,          
     \label{aeqconsBW}
\end{equation}
where 
\begin{equation}
    {\cal N}_{\rm R}^2 \equiv \rmi \, {\rm res}[S(E)]_{E=z_{\rm R}} \equiv 
          \rmi \, {\rm r}_{\rm R} \, , 
       \label{normafac}
\end{equation}
$S$ denotes the $S$ matrix, and ${\rm r}_{\rm R}$ denotes the residue
of $S$ at $z_{\rm R}$. In addition, we will see that the lower bound of 
the energy (threshold) is the reason why this amplitude is not exactly but 
only approximately given by the Breit-Wigner amplitude.

Section~\ref{sec:basics} provides a quick summary of the most important 
properties of the Gamow states, along with some basic phenomenological 
properties of resonances. The proof of~(\ref{aeqconsBW}) is provided in
Sec.~\ref{sec:proof}. The conclusions are included in 
Sec.~\ref{sec:conclusions}. 

For the sake of clarity, we shall prove Eq.~(\ref{aeqconsBW}) using the
example of the spherical shell potential for zero angular momentum. However,
as explained in Appendix~\ref{sec:appendix-gener}, the result is valid for 
any partial wave and for spherically symmetric potentials that fall off 
faster than exponentials. Finally, in Appendix~\ref{sec:appendix-cdf},
we provide a thorough characterization of the complex delta function
and its associated functional, since they have rarely appeared in the 
literature.

\section{Basics of resonances and Gamow states}
\setcounter{equation}{0}
\label{sec:basics}

Resonance peaks are characterized by the energy $E_{\rm R}$ 
at which they occur and by their width $\Gamma _{\rm R}$. The resonance
peak is related to a pole of the $S$ matrix at the complex number
$z_{\rm R}= E_{\rm R} - \rmi \, \Gamma _{\rm R}/2$, because the theoretical 
expression of the cross section in terms of the $S$ matrix fits the
experimental cross section in the neighborhood of $E_{\rm R}$, see 
Eqs.~(\ref{Smex}) and~(\ref{crossS}) below.

When the peak is too narrow and its width cannot be measured, one measures 
the lifetime $\tau _{\rm R}$ of the decaying particle. Decaying systems follow 
the exponential decay law, except for short- and long-term deviations.

Although a decaying particle has a finite lifetime, it is otherwise assigned 
all the properties that are attributed to stable particles, like
angular momentum, charge, spin and parity. For example, a radioactive nucleus 
has a finite lifetime, but otherwise it possesses all the properties of stable 
nuclei; in fact, it is included in the periodic table of the elements along 
with the stable nuclei. Similarly, most elementary particles are 
unstable, and they are listed along with the stable ones in the Particle Data 
Table~\cite{PDT} and attributed
values for the mass, spin and width (or lifetime). Thus, stable particles
differ from unstable ones by the value of their width, which is zero
in the case of stable particles and different from zero in the case of 
unstable ones. Hence, phenomenologically, unstable particles are not less
fundamental than the stable ones.

A priori, resonances and decaying particles are different entities. A 
resonance refers to the energy distribution of the outgoing particles in
a scattering process, and it is characterized by its energy and 
width. A decaying state is described in a time-dependent setting by its
energy and lifetime. Yet the difference is quantitative rather than 
qualitative, and both concepts are related by
\begin{equation}
       \Gamma _{\rm R}=\frac{\hbar}{\tau _{\rm R}} \, ,
          \label{lifewidre}
\end{equation}
though in most systems one can measure either $\tau _{\rm R}$ or
$\Gamma _{\rm R}$, but not both.

Theoretically, however, the relation~(\ref{lifewidre}) is usually justified
as an approximation, $\tau _{\rm R}\Gamma _{\rm R} \sim \hbar$, as a kind of 
time-energy uncertainty relation. For a long time, it was not possible to
experimentally check whether the relation~(\ref{lifewidre}) is exact or 
approximate, since 
the lifetime and width could not be measured in the same system. This
changed with the measurements of the width~\cite{OATES} and 
lifetime~\cite{VOLZ} of the $3p\ ^2P_{3/2}$ state of Na, which provide a firm 
experimental basis that Eq.~(\ref{lifewidre}) holds exactly, not just 
approximately. Thus, resonances and decaying systems are two sides of the 
same phenomenon.

Although the resonance peaks in the cross section resemble the Lorentzian, 
the resonance lineshape does not coincide exactly with the 
Lorentzian. Two features of the cross section reveal so. First, the maximum 
of the resonance peak never occurs at $E=E_{\rm R}$, whereas the maximum
of the Lorentzian occurs exactly at $E=E_{\rm R}$. And second, the Laurent 
expansion of the $S$ matrix around the resonance pole,
\begin{equation}
      S(E)=\frac{{\rm r}_{\rm R}}{E-z_{\rm R}}+B(E) \, , 
                  \label{Smex}
\end{equation}
which produces the Lorentzian peak in the cross section~\cite{NOTE1}, 
\begin{equation}
     \sigma \sim \frac{1}{(E-E_{\rm R})^2+(\Gamma _{\rm R}/2)^2} \, ,
            \label{crossS}
\end{equation}
is valid only in the vicinity of the resonance pole. Because~(\ref{Smex}) 
and~(\ref{crossS}) are valid only in the vicinity of the resonance energy, the 
Lorentzian lineshape is just a near-resonance approximation to the
exact resonance lineshape. 

Because the Lorentzian does not coincide exactly with the resonance lineshape,
the Breit-Wigner amplitude cannot coincide exactly with the resonance 
amplitude. One can reach the same conclusion by using the point of view of
decaying states as follows. The Breit-Wigner amplitude 
yields the exponential decay law only when it is defined over the
whole of the energy real line $(-\infty , \infty )$ rather than just over the
scattering spectrum (see e.g.~\cite{FONDA}). Because in quantum mechanics 
the scattering spectrum has a lower bound, the Breit-Wigner amplitude would 
yield the exponential decay law only if it was defined also at energies that 
do not belong to the scattering spectrum. Thus, the Breit-Wigner amplitude is 
incompatible with the exponential decay law, and therefore cannot coincide 
with the exact resonance/decay amplitude.

Mathematically, the Gamow states are eigenvectors of the Hamiltonian with 
a complex eigenvalue $z_{\rm R}=E_{\rm R} -\rmi \, \Gamma _{\rm R}/2$,
\begin{equation}
       H|z_{\rm R}\rangle = z_{\rm R} |z_{\rm R}\rangle \, ,
                \label{tiSeq} 
\end{equation}
and, in the radial position representation, they satisfy a 
``purely outgoing boundary condition'' (POBC) at infinity:
\begin{equation}
      \langle r|z_{\rm R}\rangle \sim 
             \rme ^{{\rm i}\sqrt{(2m/\hbar ^2)z_{\rm R}} \, r} 
          \, , \quad  \mbox{as} \  r \to \infty \, .
         \label{POBC}
\end{equation}
The time-independent Schr\"odinger equation~(\ref{tiSeq}) subject to the 
POBC~(\ref{POBC}) is equivalent to the following integral equation of the 
Lippmann-Schwinger type:
\begin{equation}
    |z_{\rm R}\rangle =\frac{1}{z_{\rm R}-H_0+\rmi \hskip0.2mm 0}
               V |z_{\rm R}\rangle \, ,
         \label{inteGam}
\end{equation}
where $H_0$ is the free Hamiltonian and $V$ is the potential. Since 
Eq.~(\ref{inteGam}) also yields 
the bound states, the Gamow states are a natural generalization to resonances 
of the wave functions of bound states. 
The bound and 
resonance energies obtained by solving~(\ref{inteGam}) coincide with the 
poles of the $S$ matrix. 

The time evolution of a Gamow state is given by
\begin{equation}
       \rme ^{-\rmi Ht/\hbar}|z_{\rm R}\rangle = 
        \rme ^{-\rmi E_{\rm R}t/\hbar} \rme ^{-\Gamma _{\rm R}t/(2\hbar)} 
            |z_{\rm R}\rangle  \, ,
                \label{tdSeq} 
\end{equation}
and therefore the Gamow states abide by the exponential decay law. Because
the eigenvalue of Eq.~(\ref{tiSeq}) is also a pole of the $S$ matrix, 
Eq.~(\ref{tdSeq}) implies that Eq.~(\ref{lifewidre}) holds. In this 
way, the Gamow states unify the concepts of resonance and decaying 
particle, and they provide a ``particle status'' for them.

Furthermore, since one can obtain both the bound and the resonance energies 
from Eq.~(\ref{inteGam}), or from the poles of the $S$ matrix, resonances are 
qualitatively the same as bound states. The only difference is 
quantitative: The Gamow states have a non-zero width (i.e., finite lifetime), 
whereas the bound states have a zero width (i.e., infinite lifetime). 

An important feature of the Gamow states is that they form a basis that expands
any wave packet $\varphi ^+$, see e.g.~review~\cite{05CJP}. The basis 
formed by the Gamow states is not complete though, and one has to add an 
additional set of kets to complete the basis. In a system with several 
resonances, we have that
\begin{equation}
     \varphi ^+(t)=\sum_n \rme ^{-\rmi z_nt/\hbar}
                   |z_n\rangle \langle z_n|\varphi ^+\rangle 
      +\int_0^{-\infty}\rmd E \, \rme ^{-\rmi Et/\hbar}
                 |E^+\rangle \langle ^+E|\varphi ^+ \rangle  \, ,
        \label{resoexpan}
\end{equation}
where $z_n=E_n-\rmi \, \Gamma _n/2$ denotes the $n$th resonance energy. In this
equation, the sum contains the resonance contribution, whereas the integral
contains the background. For simplicity, we have omitted the 
contribution from the bound states. The main virtue of resonance expansions 
is to isolate each resonance's contribution to the wave packet.

Resonance expansions allow us to understand the deviations from exponential 
decay~\cite{RAIZEN}. In the energy region where one resonance ${\rm R}$ is 
dominant, the expansion~(\ref{resoexpan}) can be written as
\begin{equation}
      \varphi ^+(t)= \rme ^{-\rmi z_{\rm R}t/\hbar}
            |z_{\rm R}\rangle \langle z_{\rm R}|\varphi ^+\rangle +
            {\rm background(R)} \, ,
        \label{resoexpanR}
\end{equation}
where the term ``background(R)'' contains all contributions not associated
with the resonance ${\rm R}$, including those from other resonances. Because 
``background(R)'' will always be nonzero, there will always be deviations 
from exponential decay. The magnitude of these deviations depends
on how well we tune the system around the resonance energy: The better
we tune the system around the Gamow state $|z_{\rm R}\rangle$, the smaller
``background(R)'' will be. Note that ``background(R)'' is the analog to 
the background $B(E)$ of the expansion~(\ref{Smex}).

\section{Proof}
\setcounter{equation}{0}
\label{sec:proof}

\subsection{Preliminaries}
\label{sec:prelmi}

The proof of~(\ref{aeqconsBW}) presented below is a straightforward application
of the theory of distributions. Rather than working in a general setting, we 
will use the example of the spherical shell potential,
\begin{equation}
          V({\bf x})= V(r)=\left\{ \begin{array}{ll}
                                0   &0<r<a  \\
                                V_0 &a<r<b  \\
                                0   &b<r<\infty \, ,
                  \end{array} 
                 \right. 
	\label{potential}
\end{equation}
and restrict ourselves to the s partial wave. 

In order to prove~(\ref{aeqconsBW}), we need to recall that in quantum
mechanics, the transition amplitude from one state to another is 
given by the scalar product of those states: 
\begin{equation}
     {\cal A}(z_{\rm R} \to E)= \langle ^-E|z_{\rm R}\rangle \, ,\quad 
                E\geq 0  \, ,
          \label{tramaid}
\end{equation}
where $\langle ^-E|$ is the ``out'' bra solution of the Lippmann-Schwinger 
equation. 

For the potential~(\ref{potential}), the ``out'' Lippmann-Schwinger 
eigenfunction reads, in the radial position representation, as
\begin{equation}
      \langle r|E^{-}\rangle \equiv 
       \chi ^{-}(r;E)= N(E) \,  
     \frac{\chi (r;E)}{{\cal J}_{-}(E)}\, , \qquad E\in [0,\infty ) \, ,
      \label{pmeigndu}
\end{equation}
where $N(E)$ is a delta-normalization factor,
\begin{equation}
      N(E)=\sqrt{\frac{1}{\pi}
         \frac{2m/\hbar ^2}{\sqrt{2m/\hbar ^2 \, E \,}\,}\,} \, ,
        \label{Nfactor}
\end{equation}
$\chi (r;E)$ is the regular solution,
\begin{equation}
       \chi (r;E) = \left\{
             \begin{array}{ll}
                  \sin (kr)  & 0<r<a \\
               {\cal J}_1(E)\rme ^{\rmi Qr}
              +{\cal J}_2(E)\rme ^{-\rmi Qr} & a<r<b \\ 
                  {\cal J}_3(E) \rme ^{\rmi kr}
                   +{\cal J}_4(E) \rme ^{-\rmi kr}
                                                         & b<r<\infty \, ,
               \end{array}
              \right.
        \label{LSchi}
\end{equation}
the wave numbers $k$ and $Q$ are given by
\begin{equation}
    k=\sqrt{ \frac{2m}{\hbar^2}E\,} \, , \quad \, 
    Q=\sqrt{\frac{2m}{\hbar ^2}(E-V_0)\,} \, ,
\end{equation}
and ${\cal J}_{\pm}(E)$ are the Jost functions,
\begin{equation}
        {\cal J}_{+}(E)= -2\rmi {\cal J}_4(E) \, , \quad  \,
        {\cal J}_-(E)=2\rmi {\cal J}_3(E) \, .
\end{equation}
The resonance energies $z_{\rm R}$ produced by the potential~(\ref{potential}) 
coincide with the zeros of ${\cal J}_{+}$. With each resonance 
energy $z_{\rm R}$, we associate a Gamow eigenfunction $u(r;z_{\rm R})$:
\begin{equation}
  \langle r|z_{\rm R} \rangle =u(r;z_{\rm R})= 
            \sqrt{\frac{m}{\hbar ^2 k_{\rm R}}\,} {\cal N}_{\rm R} \times
         \left\{ \begin{array}{ll}
         \frac{1}{{\mathcal J}_3(k_{\rm R})}\sin(k_{\rm R}r)  &0<r<a \\ [1ex]
         \frac{{\mathcal J}_1(k_{\rm R})}{{\mathcal J}_3(k_{\rm R})} \,
                \rme ^{\rmi Q_{\rm R}r}
         +\frac{{\mathcal J}_2(k_{\rm R})}{{\mathcal J}_3(k_{\rm R})} \,
                \rme ^{-\rmi Q_{\rm R}r} &a<r<b 
         \\  [1ex]
         \rme ^{\rmi k_{\rm R}r}  &b<r<\infty \, ,
                           \end{array} 
                  \right.
	\label{dgv0p} 
\end{equation}
where
\begin{equation}
       k_{\rm R}= \sqrt{\frac{2m}{\hbar ^2}z_{\rm R}\,} \, , \quad 
       Q_{\rm R}= \sqrt{\frac{2m}{\hbar ^2}(z_{\rm R}-V_0)\,} \, ,
\end{equation}
and where ${\cal N}_{\rm R}$ is given by Eq.~(\ref{normafac}). From 
Eqs.~(\ref{pmeigndu}) and (\ref{dgv0p}) we obtain
\begin{equation}
        \chi ^-(r;z_{\rm R}) = \frac{1}{\rmi\sqrt{2\pi}{\cal N}_{\rm R}} 
            u(r;z_{\rm R}) \, .
            \label{relaniodripp}
\end{equation}

As shown in~\cite{LS2}, the analytic continuations of the Lippmann-Schwinger
kets --and therefore also the Gamow kets-- are well defined as antilinear 
functionals over the space of test functions $\psi ^-$ for which the 
following quantities are finite: 
\begin{equation}
   \| \psi ^{-}\|_{n,n'} := \sqrt{\int_{0}^{\infty}\rmd r \, 
    \left| \frac{nr}{1+nr}\, \rme ^{nr^2/2} (1+H)^{n'}
             \psi ^{-}(r) \right|^2 \, } 
                 \, , \quad n,n'=0,1,2, \ldots  
      \label{normsLS}
\end{equation}
The action of the Gamow ket $|z_{\rm R}\rangle$ on the test functions 
$\psi ^-$ is explicitly given by
\begin{equation}
      \langle \psi ^-|z_{\rm R} \rangle =
          \int_0^{\infty}\rmd r \, \langle \psi ^-|r \rangle
             \langle r|z_{\rm R} \rangle =
          \int_0^{\infty}\rmd r \, \psi ^-(r)^* u(r;z_{\rm R})  \, .
            \label{wendned}
\end{equation}

For the sake of clarity, we need to introduce a special notation that will 
specify when we are working in the energy representation: Whenever we work 
in such representation, we will add a hat to the corresponding quantity. For 
example, the energy representation of a wave function $\psi ^-(r)$ will be 
denoted by $\widehat{\psi}^-(E)$.

By using the operator $U_-$ of~\cite{LS1}, one can obtain the energy 
representation of $\psi ^-(r)$,
\begin{equation}
     \widehat{\psi}^-(E)=(U_-\psi ^-)(E)=
       \int_0^{\infty}\rmd r \, \psi ^-(r) \chi ^-(r;E)^*  \, .
      \label{-LSinteexpre}
\end{equation}
As shown in~\cite{LS2}, when
$\psi ^-(r)$ satisfies~(\ref{normsLS}), the analytic continuation of
$[\widehat{\psi}^-(E)]^*$ exists. We shall denote such analytic continuation
by $\widehat{\psi}^-(z^*)^*$:
\begin{equation}
     \widehat{\psi}^-(z^*)^*=
       \int_0^{\infty}\rmd r \, \psi ^-(r)^* \chi ^-(r;z)  \, .
      \label{anaconcoc}
\end{equation}
Not only $\widehat{\psi}^-(z^*)^*$ exist,
all the test functions $[\widehat{\psi}^-(z^*)]^*$ are analytic
at the resonance energies.\footnote{The poles of
$[\widehat{\psi}^-(z^*)]^*$ are located on the first sheet of the Riemann
surface, that is, on the upper half of the wave-number plane.} As 
explained in Appendix~\ref{sec:appendix-cdf}, this means that the 
antilinear complex delta functional at the resonance energies $z_{\rm R}$
can be defined as
\begin{equation}
    \begin{array}{rcl}
       \widehat{\delta}_{z_{\rm R}} : \widehat{\Phi}_{-\rm exp} & 
                 \longmapsto & 
                                 {\mathbb C}  \\
       \widehat{\psi}^- & \longmapsto & 
        \langle \widehat{\psi}^-|\widehat{\delta}_{z_{\rm R}} \rangle \equiv  
  \widehat{\delta}_{z_{\rm R}}(\widehat{\psi}^-) :=
                    \widehat{\psi}^- (z_{\rm R}^*)^*  \, ,
    \end{array}
        \label{Cdelta-anti-mt}
\end{equation}
where $\widehat{\Phi}_{-\rm exp}$ is the space of test functions
$\widehat{\psi}^-$. That is, $\widehat{\delta}_{z_{\rm R}}$ associates
a test function $\widehat{\psi}^-$ with the value that the analytic 
continuation of $[\widehat{\psi}^-]^*$ takes at $z_{\rm R}$.

\subsection{The Gamow state and the complex delta function}
\label{sec:GSandcdf}

In order to obtain the equality of Eq.~(\ref{aeqconsBW}), we are going first
to denote the energy representation of the Gamow ket as
\begin{equation}
      |\widehat{z}_{\rm R}^{-}\rangle \equiv 
       U_{-}|z_{\rm R}\rangle \, .
\end{equation}
Then, it follows that
\begin{eqnarray}
      \langle \widehat{\psi}^-|\widehat{z}_{\rm R}^-\rangle &=&
       \langle \widehat{\psi}^-|U_-|z_{\rm R} \rangle \nonumber  \\
       &=& \langle U_-^{\dagger}\widehat{\psi}^-|z_{\rm R} \rangle \nonumber \\
       &=& \langle \psi ^-|z_{\rm R} \rangle \nonumber \\
       &=& \int_0^{\infty}\rmd r \, [\psi ^-(r)]^* u(r;z_{\rm R}) 
          \hskip2.97cm  \mbox{by~(\ref{wendned})}  \nonumber \\
       &=& \rmi \sqrt{2\pi \, } \,  {\cal N}_{\rm R}  \int_0^{\infty}\rmd r \, 
            \psi ^-(r)^* \chi ^-(r;z_{\rm R})
          \hskip1.20cm  \mbox{by~(\ref{relaniodripp})}  \nonumber \\
       &=& \rmi \sqrt{2\pi \, } \,  {\cal N}_{\rm R}  \, 
           \widehat{\psi}^-(z_{\rm R}^*)^* 
               \hskip3.94cm  \mbox{by~(\ref{anaconcoc})} \nonumber  \\
       &=& \rmi \sqrt{2\pi \, } \,  {\cal N}_{\rm R} 
           \langle \widehat{\psi}^-|\widehat{\delta}_{z_{\rm R}} \rangle 
              \hskip3.96cm       \mbox{by~(\ref{Cdelta-anti-mt})} \, . 
       \label{profosjd}
\end{eqnarray}
This equation proves that the energy representation of the Gamow functional,
$|\widehat{z}_{\rm R}^- \rangle$, is proportional to the antilinear complex
delta functional, $|\widehat{\delta}_{z_{\rm R}} \rangle$. 

In the energy representation, the 
identity $\widehat{\mathbb I}$ can be written as
\begin{equation}
    \int_0^{\infty}\rmd E \, |\widehat{E}^-\rangle \langle ^-\widehat{E}| =
          \widehat{\mathbb I} \, ,
     \label{identity}
\end{equation}
where $|\widehat{E}^-\rangle$ denotes the energy representation of
$|E^-\rangle$. By inserting~(\ref{identity}) into the first and the last 
terms of~(\ref{profosjd}), we obtain
\begin{equation}
    \int_0^{\infty}\rmd E \, \langle \widehat{\psi}^-|\widehat{E}^-\rangle 
          \langle ^-\widehat{E}|\widehat{z}_{\rm R}^- \rangle =
 \int_0^{\infty}\rmd E \, \rmi \sqrt{2\pi \, } {\cal N}_{\rm R} 
    \langle \widehat{\psi}^-|\widehat{E}^-\rangle 
          \langle ^-\widehat{E}|\widehat{\delta}_{z_{\rm R}}\rangle  \, .
     \label{identity2}
\end{equation}
Because Eq.~(\ref{identity2}) holds for any $\widehat{\psi}^-$, it follows
that
\begin{equation}
    \langle ^-\widehat{E}|\widehat{z}_{\rm R}^- \rangle =
              \rmi \sqrt{2\pi \, } {\cal N}_{\rm R} 
    \langle ^-\widehat{E}|\widehat{\delta}_{z_{\rm R}}\rangle \equiv 
       \rmi \sqrt{2\pi \, } {\cal N}_{\rm R} \, \delta(E-z_{\rm R}) \, ,
     \label{identity3}
\end{equation}
which, after dropping the hat notation and using Eq.~(\ref{tramaid}),
becomes the equality in Eq.~(\ref{aeqconsBW}).

\subsection{The Gamow state and the Breit-Wigner amplitude}
\label{sec:GSandBW}

In order to obtain the approximation of Eq.~(\ref{aeqconsBW}),
we are going to obtain first the transition 
amplitude $\widetilde{ {\cal A}}(z_{\rm R} \to E)$ from a resonance 
state of energy $z_{\rm R}$ to a scattering state of energy 
$-\infty < E< \infty$,
\begin{equation}
     \widetilde{ {\cal A}}(z_{\rm R} \to E)=
      \langle ^-E|z_{\rm R}\rangle \, , \quad E\in (-\infty , \infty ) \, .
        \label{tildeamp}
\end{equation}
Even though a quantum system can only decay to a scattering state of energy 
$E\geq 0$, we are going to ask the system to pretend that 
it can also decay to negative energies. Mathematically, this is equivalent
to ask the system to pretend that its scattering spectrum runs from $-\infty$ 
to $+\infty$. Physically, it is equivalent to ignore the effect of the
lower bound of energy $E=0$. Calculating~(\ref{tildeamp}), i.e.,
forcing the system to decay also to negative energies, needs a regulator. The 
regulator we will use is $\rme^{-\rmi \alpha E}$, where $\alpha >0$ and
$E$ has zero or negative imaginary part. The reason why we use this regulator
is that for complex $z$, the wave functions $\widehat{\psi}^-(z^*)^*$ grow 
slower than $\rme ^{|{\rm Im}(\sqrt{2m/\hbar ^2 \, z})|^2}$ in the lower half 
plane of the second sheet~\cite{LS2}. More precisely, Proposition~3 
in~\cite{LS2} shows that for each $n=1,2,\ldots$ and for each $\beta >0$,
there is a $C>0$ such that in the lower half plane 
of the second sheet, $\widehat{\psi}^-(z^*)^*$ is bounded by
\begin{equation}
    |\widehat{\psi}^-(z^*)^*| \leq C \, 
      \frac{1} {|z|^{1/4}|1+z|^{n}} \,
     \rme ^{\frac{\, \, |{\rm Im}(\sqrt{2m/\hbar ^2 \, z\,})|^2}{2\beta}} \, .
    \label{boundinwhvalp+}
\end{equation}
This estimate implies that $\rme^{-\rmi\alpha z} \widehat{\psi}^-(z^*)^*$ 
tends to zero in the infinite arc of the lower half of the second sheet, 
\begin{equation}
     \lim_{z\to \infty} \rme ^{-\rmi z\alpha}\widehat{\psi}^-(z^*)^* = 0 \, ,
         \quad \alpha >0 \, .
        \label{indifli}
\end{equation}
In its turn, the limit~(\ref{indifli}) enables us to apply Cauchy's theorem 
to obtain
\begin{equation}
    \widehat{\psi}^-(z_{\rm R}^*)^*=
    \lim _{\alpha \to 0}- \frac{1}{2\pi \rmi} \int_{-\infty}^{\infty}\rmd E \,
     \rme ^{-\rmi \alpha E} \widehat{\psi}^-(E)^* \frac{1}{E-z_{\rm R}} \, ,
         \label{CAUSK}
\end{equation}
where the integral is performed infinitesimally below the real axis of the
second sheet. By 
multiplying both sides of~(\ref{CAUSK}) by $\rmi \sqrt{2\pi}{\cal N}_{\rm R}$,
and by recalling~(\ref{profosjd}), we obtain
\begin{equation}
     \langle \widehat{\psi}^-|\widehat{z}_{\rm R}^-\rangle  =
    \lim _{\alpha \to 0} \int_{-\infty}^{\infty}\rmd E \,
     \rme ^{-\rmi\alpha E} \psi ^-(E)^* (-1)
          \frac{{\cal N}_{\rm R}}{\sqrt{2\pi}} \frac{1}{E-z_{\rm R}} \, .
            \label{itenrmind}
\end{equation}
In the bra-ket notation, Eq.~(\ref{itenrmind}) reads as 
\begin{equation}
    \langle \widehat{\psi}^-|\widehat{z}_{\rm R}\rangle =
    \lim _{\alpha \to 0} \int_{-\infty}^{\infty}\rmd E \,
     \rme ^{-\rmi\alpha E} \langle \widehat{\psi}^-|\widehat{E}^-\rangle 
           \langle ^-\widehat{E}|\widehat{z}_{\rm R}\rangle  \, . 
      \label{itenrmindb-k}
\end{equation}
Comparison of~(\ref{itenrmind}) with~(\ref{itenrmindb-k}) yields the
following expression for the amplitude~(\ref{tildeamp}):
\begin{equation}
     \widetilde{ {\cal A}}(z_{\rm R} \to E) =
           -\frac{{\cal N}_{\rm R}}{\sqrt{2\pi}} \frac{1}{E-z_{\rm R}} \, , 
           \ E\in (-\infty , \infty ) \, .
\end{equation}
Thus, if the scattering spectrum was the whole real line, the resonance
amplitude would be exactly the Breit-Wigner amplitude. However, because
the scattering spectrum has a lower bound, the resonance amplitude is not 
exactly the Breit-Wigner amplitude. Only when we can neglect the effect
of the threshold, the resonance amplitude coincides with the Breit-Wigner
amplitude:
\begin{equation}
    {\cal A}(z_{\rm R} \to E) \simeq \widetilde{ {\cal A}}(z_{\rm R} \to E) 
             \, ,
\end{equation}
which is the approximation on the right-hand side of~(\ref{aeqconsBW}). In
particular, when the threshold can be ignored, the complex delta function 
becomes for all intends and purposes the Breit-Wigner amplitude.

It should be stressed that the amplitude~(\ref{tildeamp}) is not physical, 
because in~(\ref{tildeamp}) the energy $E$ runs over the whole real line rather
than over the scattering spectrum. However, such unphysical amplitude helps us 
understand what the physical amplitude --the complex delta function-- is, by 
allowing us to see how the resonance would decay if the scattering 
spectrum was the whole real line.

\subsection{Further remarks}
\label{sec:fR}

Aside from phase space factors, cross sections are determined by the 
transition amplitude from an ``in'' to an ``out'' state,
${\cal A}(E_i\to E_f )$. If $|E^{\pm}\rangle$ denote the ``in'' and ``out''
solutions of the Lippmann-Schwinger equation, then 
\begin{equation}
       {\cal A}(E_i\to E_f ) = \langle ^-E_f|E_i^+\rangle =
          S(E_i) \delta (E_f -E_i) \, .
\end{equation}
If we imagine now that instead of an initial state $|E_i^+\rangle$ we had an 
unstable particle $|z_{\rm R}\rangle$, the transition (decay) amplitude 
${\cal A}(z_{\rm R} \to E_f )$ would be given by~(\ref{aeqconsBW}). Using the
approximate decay amplitude of~(\ref{aeqconsBW}), one obtains the following
approximate decay probability:
\begin{equation}
     |{\cal A}(z_{\rm R} \to E_f )|^2 \simeq
    \frac{|{\cal N}_{\rm R}|^2}{2\pi} 
    \frac{1}{(E_f-E_{\rm R})^2+ (\Gamma _{\rm R}/2)^2} \, ;  
        \label{almoslpro}
\end{equation} 
that is, the decay probability of a resonance is given by the Lorentzian
when the effect of the threshold can be ignored.\footnote{A much more
detailed study of the dependence of the cross section (and
expectation values of observables) on the Breit-Wigner
amplitude can be found in e.g.~\cite{SIEGERT,BERGGREN78,BERGGREN96}.} Because 
the almost-Lorentzian 
decay probability~(\ref{almoslpro}) coincides with the almost-Lorentzian
peaks in cross sections, resonances can be interpreted as intermediate, 
unstable particles. 

Finally, it is worthwhile to compare the Gamow states with the states 
introduced by Kapur and Peierls~\cite{KAPUR}. As mentioned above, the
Gamow states are eigenfunctions of the Hamiltonian that satisfy the 
POBC~(\ref{POBC}) at infinity; the wave numbers involved in the 
POBC~(\ref{POBC}) are complex and proportional to the square root of the
complex eigenenergies of the Gamow states; such complex eigenenergies are the 
same as the poles of the $S$ matrix, and they do not depend on any external 
parameter or energy. By contrast, the Kapur-Peierls states are 
eigenfunctions of the Hamiltonian that satisfy a POBC at a finite radial 
distance $r_0$, where $r_0$ is such that the potential vanishes for $r>r_0$; 
the wave numbers involved in the POBC satisfied by the Kapur-Peierls 
states are real and proportional to the square root of the real energy
of the incoming particle; the POBC satisfied by the Kapur-Peierls states 
makes them 
and their associated complex eigenenergies depend on $r_0$ and on the real 
energy of the incoming particle; also, the complex eigenenergies of the 
Kapur-Peierls states are not the same as the poles of the $S$ matrix. Thus, the 
Kapur-Peierls states do not seem to be related to the standard Breit-Wigner 
amplitude, because such amplitude does not depend on $r_0$ and its complex 
energy does not depend on the energy of the incoming particle.

\section{Conclusions}
\label{sec:conclusions}

Since resonances leave an almost-Lorentzian fingerprint in the cross section,
and since the Gamow states are the wave functions of resonances, the decay
amplitude provided by a Gamow state should be linked to the Breit-Wigner
amplitude. In this paper, we have found that the precise link is given by
Eq.~(\ref{aeqconsBW}), and we have interpreted this result by saying that
the resonance amplitude associated with a Gamow state is exactly given by
the complex delta function, and that the Breit-Wigner amplitude is an 
approximation to such resonance amplitude, which approximation is valid when 
we can neglect the effect of the threshold. Thus, Eq.~(\ref{aeqconsBW})
establishes the precise relation between the Gamow state, Nakanishi's 
complex delta function and the Breit-Wigner amplitude. In addition, 
Eq.~(\ref{aeqconsBW}) affords another theoretical argument in favor of
interpreting the almost-Lorentzian peaks in cross sections as intermediate, 
unstable particles---resonances are real (as opposed to virtual) particles, 
in accordance with resonance phenomenology. 

As is well known, the actual resonance lineshape of cross sections can be 
very different from a quasi-Lorentzian one, due to the effect of thresholds, 
other resonances, or extra channels. The usefulness 
of~(\ref{aeqconsBW}) does not lie in predicting the exact shape of the cross 
section, but rather in identifying what contribution to the cross section 
comes from each pole of the $S$ matrix. In particular, although the
equality in Eq.~(\ref{aeqconsBW}) is always exact, for practical purposes 
the approximation in Eq.~(\ref{aeqconsBW}) is useful only for narrow resonances.

When we add Eq.~(\ref{aeqconsBW}) to the other known properties of the Gamow 
states, we see that such states have all the necessary properties to describe 
resonance/unstable particles:
\begin{itemize}
   \item[$\bullet$] They are associated with poles of the $S$ matrix. 

   \item[$\bullet$]They exhibit the correct phenomenological signatures of 
both resonances (almost-Lorentzian lineshape) and unstable particles 
(exponential decay), and they provide a firm theoretical basis 
for~(\ref{lifewidre}).

   \item[$\bullet$] They are basis vectors that isolate each resonance's 
contribution to a wave packet.
\end{itemize}

\section*{Acknowledgments}

The author thanks Casey Koeninger for encouragement, and Alfonso Mondrag\'on 
for enlightening criticisms. The author thanks Alfonso Mondrag\'on also for
his precise explanation of the relation between the Gamow and the Kapur-Peierls
states.

\appendix
\def\thesection{\Alph{section}}
\section{Generalizations}
\setcounter{equation}{0}
\label{sec:appendix-gener}

Equation~(\ref{aeqconsBW}) is not valid only for the spherical
shell potential~(\ref{potential}) but actually holds for a quite large
class of potentials. The reason can be found in well-known results of 
scattering 
theory~\cite{TAYLOR,NUSSENZVEIG}. As explained in~\cite{TAYLOR}, page~191,
partial wave analysis is valid whenever the spherically symmetric potential 
satisfies the following requirements:
\begin{enumerate}
    \item[$\widetilde{\rm I}$.] $V(r)=O(r^{-3-\epsilon})$ as $r\to \infty$.
    \item[II.] $V(r)=O(r^{-3/2+\epsilon})$ as $r\to 0$.
     \item[III.] $V(r)$ is continuous for $0<r<\infty$, except perhaps at a 
finite number of finite discontinuities.
\end{enumerate}
These conditions are, however, not sufficient to guarantee that the $S$ matrix
$S(E)$, the Jost functions ${\cal J}_{\pm}(E)$ and the Lippmann-Schwinger 
eigenfunction $\chi ^{-}(r;E)$ can be analytically continued into the whole 
complex plane. Such analytic continuation is guaranteed when we replace 
condition~$\widetilde{\rm I}$ by the more stringent
\begin{enumerate}
     \item[I.] $V(r)$ falls off faster than exponentials as $r\to \infty$,
\end{enumerate}
as stated throughout Chapters 11 and 12 of~\cite{TAYLOR}, and in Chapter~5 
of~\cite{NUSSENZVEIG}, especially in 
Theorem~5.3.2. Thus, when $V(r)$ satisfies I-III, even though we may not know 
their exact analytic expressions, we know that $S(E)$, ${\cal J}_{\pm}(E)$ and 
$\chi ^{-}(r;E)$ can be analytically continued into the whole complex plane 
and that the Gamow eigenfunction $u(r;z_{\rm R})$ is well defined. Moreover, 
since in the asymptotic region $r\to \infty$ the expressions of 
$u(r;z_{\rm R})$ and $\chi ^-(r;E)$ for any potential satisfying I-III are 
the same as the expressions of $u(r;z_{\rm R})$ and $\chi ^-(r;E)$ for the 
spherical shell potential in the region $r>b$ (with different expressions 
for the Jost functions), the general proof goes through exactly the same lines 
as the proof for the spherical shell potential. Finally, the argument extends 
without difficulty to higher angular momentum.

\def\thesection{\Alph{section}}
\section{The complex delta functional}
\setcounter{equation}{0}
\label{sec:appendix-cdf}

In quantum mechanics, the complex delta function was originally introduced by
Nakanishi~\cite{NAKANISHI} to describe resonances in the Lee 
model~\cite{LEE}. In mathematics, the complex delta function was introduced
by Gelfand and Shilov~\cite{GELFAND}. The purpose of this appendix is to 
introduce the precise mathematical definition of the complex delta 
function and to show that, when the test functions are analytic, such 
definition coincides with the one given by Nakanishi.

\subsection{Three definitions of the (linear) complex delta functional}

The complex delta functional has different forms depending on the properties
of the test functions on which it acts. We shall review the three most
important forms, namely when the complex delta functional acts on 
analytic functions (this form is used in this paper and and was
introduced in~\cite{GELFAND}), when it acts on meromorphic functions (this is 
the form used by Nakanishi~\cite{NAKANISHI}), and when
it acts on non-meromorphic functions (this form was introduced 
in~\cite{GELFAND}). When the space of test functions are analytic, 
as is our case, these three forms coincide (as they should) and can be written 
as in Eq.~(\ref{Cdelta-anti-mt}).

\subsubsection{First definition---the test functions are analytic}
\label{sec:firsdief}

According to page~1 of Volume~I of Ref.~\cite{GELFAND}, a distribution is a 
function that associates a complex number with each function belonging to
a vector space: 
\begin{equation}
    \begin{array}{rcl}
       {\rm distribution} :\{ {\rm Space \ of \ functions}\} & \longmapsto & 
                           {\mathbb C}
                   \\
      {\rm function} & \longmapsto & {\rm complex \  number}  \, .
    \end{array}
        \label{distributions}
\end{equation}
The functions in the ``$\{ {\rm Space\ of\ functions}\}$'' are usually called 
test 
functions. Because a distribution maps functions into complex numbers, they
are usually called functionals. Such functionals can be linear or antilinear.

A more precise definition is the following. If $\Phi$ is a vector space
of test functions endowed with a topology, a linear (antilinear)
distribution $F$ is a function from $\Phi$ to $\mathbb C$
\begin{equation}
    \begin{array}{rcl}
       F : \Phi  & \longmapsto & {\mathbb C}  \\
         \phi & \longmapsto & F(\phi)  
    \end{array}
        \label{F}
\end{equation}
such that
\begin{itemize}
    \item[({\it i})] $F$ is well defined,
    \item[({\it ii})] $F$ is linear (antilinear),
    \item[({\it iii})] $F$ is continuous.
\end{itemize}

A very important example of distribution is the (linear) Schwartz delta 
functional at a real number $E$. Such functional associates 
with each test function $\phi$ the value that $\phi$ takes at $E$:
\begin{equation}
    \begin{array}{rcl}
       \delta _E : \Phi_{\rm Schw} & \longmapsto & {\mathbb C}  \\
       \phi & \longmapsto & \delta _E(\phi)= \phi (E)  \, ,
    \end{array}
        \label{SDdelta}
\end{equation}
where the test functions of $\Phi _{\rm Schw}$ are infinitely differentiable
and of polynomial falloff. It is 
straightforward to show that definition~(\ref{SDdelta}) satisfies the
above requirements~({\it i})-({\it iii}). 

The (linear) complex delta functional is defined in a completely
analogous way. As stated by Gelfand and Shilov~\cite[Vol.~2, page~85]{GELFAND},
the point $E$ in Eq.~(\ref{SDdelta}) may be complex in the spaces 
of analytic functions. If
$\Phi _{\rm anal}$ denotes a vector space of {\it analytic} functions
at the complex point $z_0$, then the linear complex delta 
functional at $z_0$ is defined as a function that associates with each test 
function $\phi$ the value that the analytic continuation of $\phi$ takes 
at $z_0$:
\begin{equation}
    \begin{array}{rcl}
       \delta _{z_0} : \Phi _{\rm anal} & \longmapsto & {\mathbb C}  \\
       \phi & \longmapsto & \delta _{z_0}(\phi)= 
                    \phi (z_0)  \, .
    \end{array}
        \label{Cdelta}
\end{equation}
Two important comments are in order here. First, the test functions of
$\Phi _{\rm anal}$ must be analytic at $z_0$; that is, $z_0$ is not
a singularity (e.g., a pole) of any $\phi$, otherwise 
definition~(\ref{Cdelta}) makes no sense. And second, the complex delta 
functional is completely specified by Eq.~(\ref{Cdelta}) because the 
test functions are analytic at $z_0$, and therefore one does not need to 
introduce any contour in the definition of~(\ref{Cdelta}), even though
one could use such a contour, as in Eq.~(\ref{Cdeltameroph}) below.

Definition~(\ref{Cdelta}) actually fulfills the 
requirements~({\it i})-({\it iii}).\footnote{In this paper, we omit any 
explicit discussion on the continuity requirement~({\it iii}). The 
reason is that first, the continuity of the complex delta function is 
guaranteed by the results of~\cite{LS2}, and second, continuity is not
essential to our main discussion.} The only property that is conceptually
challenging is~({\it i}). Because we are assuming that the test
functions are analytic at $z_0$, $\phi (z_0)$ exists and is unique, which
grants requirement~({\it i}). Thus, definition~(\ref{Cdelta}) completely,
rigorously and unambiguously defines the complex delta functional.

\subsubsection{Second definition---the test functions are meromorphic}
\label{sec:second-linear}

Many functions are not analytic
but just meromorphic. That is, when we analytically continue them, they have
isolated singularities (``poles'') in the complex plane. At such poles,
definition~(\ref{Cdelta}) makes no sense, and one has to extend it. If
$\Phi _{\rm mero}$ is a vector space of meromorphic functions at $z_0$, the 
(linear) complex delta functional at $z_0$ is defined as
\begin{equation}
    \begin{array}{rcl}
       \delta _{z_0} : \Phi _{\rm mero} & \longmapsto & {\mathbb C}  \\
       \phi & \longmapsto & \delta _{z_0}(\phi)= 
        \frac{1}{2\pi \rmi} \oint \rmd z \, \frac{\phi (z)}{z-z_0}  \, .
    \end{array}
        \label{Cdeltameroph}
\end{equation}
One can again check very easily that definition~(\ref{Cdeltameroph})
satisfies requirements~({\it i})-({\it iii}). Note that because in 
definition~(\ref{Cdeltameroph}) the test functions are 
meromorphic, such definition depends on Cauchy's theorem and on the 
contour used.\footnote{The contour used in Eq.~(\ref{Cdeltameroph}) is assumed 
to be a circle around $z_0$ such that the test function $\phi$ is analytic 
inside such circle except perhaps at $z_0$.} 

If we denote by $a_0$ the zeroth term of the Laurent expansion of
$\phi (z)$ around $z_0$, then definition~(\ref{Cdeltameroph}) associates
$a_0$ with each test function $\phi$, since
\begin{equation}
     a_0=\frac{1}{2\pi \rmi} \oint \rmd z \, \frac{\phi (z)}{z-z_0} \, . 
\end{equation}
Thus, we may write definition~(\ref{Cdeltameroph}) as
\begin{equation}
    \begin{array}{rcl}
       \delta _{z_0} : \Phi _{\rm mero} & \longmapsto & {\mathbb C}  \\
       \phi & \longmapsto &  \delta _{z_0}(\phi)= a_0  \, .
    \end{array}
        \label{Cdeltameroph-alter}
\end{equation}
Obviously, both~(\ref{Cdeltameroph}) and~(\ref{Cdeltameroph-alter}) define
the same functional, because both associate the same complex number with
the same function, even though in~(\ref{Cdeltameroph-alter}) no contour 
integral has been explicitly used.

Now, when $\phi (z)$ is analytic at $z_0$, $a_0$ is simply $\phi (z_0)$. Thus, 
when the test functions are not just meromorphic but also {\it analytic} 
at $z_0$, definitions~(\ref{Cdeltameroph}) and~(\ref{Cdeltameroph-alter}) 
become definition~(\ref{Cdelta}), {\sf because in such case all these 
definitions associate each function $\phi$ with one and the 
same complex number $\phi (z_0)$}. This is why, when the test functions
$\phi$ are all {\it analytic} at $z_0$, one can define the complex delta
functional by way of Eq.~(\ref{Cdelta}), as Gelfand and Shilov do
in page~85, Vol.~II of~\cite{GELFAND}.

\subsubsection{Third definition---the test functions are not meromorphic}
\label{sec:third-linear}

When the test functions are not meromorphic, definitions~(\ref{Cdelta}),
(\ref{Cdeltameroph}) and~(\ref{Cdeltameroph-alter}) make no sense. One 
can still define a complex delta
functional at the origin following the prescription of Gelfand and
Shilov~\cite[Vol.~I, Appendix B]{GELFAND}. When the functions are 
meromorphic, such definition of the complex delta functional at
the origin becomes~(\ref{Cdeltameroph}) and~(\ref{Cdeltameroph-alter}). 

However, because in this paper we use test functions that are analytic
at the resonance energies, we do not need to use this general definition or
definition~(\ref{Cdeltameroph}), because all these definitions actually 
become~(\ref{Cdelta}).

\subsection{Three definitions of the (antilinear) complex delta functional}
\label{sec:anti}

In this paper, we have used antilinear (rather than linear)
functionals. We will therefore briefly explain how one
defines such functionals for the cases considered in the 
previous section. 

The (antilinear) Schwartz delta functional at a real number $E$
associates with each test function $\phi$, the complex conjugate of the value 
that $\phi$ takes at $E$:
\begin{equation}
    \begin{array}{rcl}
       \widehat{\delta}_E : \Phi_{\rm Schw} & \longmapsto & {\mathbb C}  \\
       \phi & \longmapsto & \widehat{\delta}_E(\phi)= \phi (E)^*  \, .
    \end{array}
        \label{SDdelta-anti}
\end{equation}
When we write the action of $\widehat{\delta}_E$ as an integral operator, the
kernel of such integral operator is Dirac's delta function:
\begin{equation}
    \widehat{\delta}_E(\phi )= \int_0^{\infty}\rmd E' \, 
                        \delta (E'-E) \phi (E')^*
                     =\phi (E)^* \, .
        \label{SDdeltaIO}
\end{equation}

If $\Phi _{\rm anal}$ denotes a vector space of test functions $\phi$ such that
$\phi ^*$ are all {\it analytic} at $z_0$, then 
the antilinear complex delta functional at $z_0$ is a
function that associates with each test function $\phi$, the value that
the analytic continuation of $\phi ^*$ takes at $z_0$:
\begin{equation}
    \begin{array}{rcl}
       \widehat{\delta}_{z_0} : \Phi _{\rm anal} & \longmapsto & 
                                 {\mathbb C}  \\
       \phi & \longmapsto & \widehat{\delta}_{z_0}(\phi)= 
                    \phi (z_0^*)^*  \, .
    \end{array}
        \label{Cdelta-anti}
\end{equation}
When we write the expression for $\widehat{\delta}_{z_0}$ as an integral 
operator, the kernel of such integral operator is the complex delta function:
\begin{equation}
    \widehat{\delta}_{z_0}(\phi )= \int_0^{\infty}\rmd E' \, 
                   \delta (E'-{z_0}) \phi (E')^* = 
               \phi (z_0^*)^*\, .
        \label{decdf}
\end{equation}

When the test functions are only meromorphic and $z_0$ is one of their poles,
definition~(\ref{Cdelta-anti}) needs to be changed to
\begin{equation}
    \begin{array}{rcl}
     \widehat{\delta}_{z_0} : \Phi _{\rm mero} & \longmapsto & {\mathbb C}  \\
       \phi & \longmapsto & \widehat{\delta}_{z_0}(\phi)= 
        \frac{1}{2\pi \rmi} \oint \rmd z \, \frac{\phi (z^*)^*}{z-z_0}  \, .
    \end{array}
        \label{Cdeltameroph-anti}
\end{equation}
If we denote by $a_0^*$ the zeroth term of the Laurent expansion of
$\phi (z^*)^*$ around $z_0$, then definition~(\ref{Cdeltameroph-anti}) 
associates $a_0^*$ with each test function $\phi$, and therefore we can write
\begin{equation}
    \begin{array}{rcl}
     \widehat{\delta}_{z_0} : \Phi _{\rm mero} & \longmapsto & {\mathbb C} \\
       \phi & \longmapsto & \widehat{\delta}_{z_0}(\phi)= a_0^* \, .
    \end{array}
        \label{Cdeltameroph-anti-a8}
\end{equation}

If the functions are not even meromorphic, we need to use the prescription of 
Gelfand and Shilov~\cite[Vol.~I, Appendix B]{GELFAND}.

The same conclusions as in the previous section apply to the antilinear
complex delta functional. When $\phi (z^*)^*$ are all {\it analytic} 
at $z_0$, $a_0^*$ is simply $\phi (z_0^*)^*$. Thus, when the test functions 
are all {\it analytic} at $z_0$, definition~(\ref{Cdeltameroph-anti}) becomes 
definition~(\ref{Cdelta-anti}), and we are allowed to use~(\ref{Cdelta-anti}).

\subsection{Nakanishi's definition}
\label{sec:Nakadef}

Nakanishi~\cite{NAKANISHI} uses a slightly different version of the
complex delta function. When he writes $\delta _{\rm N}(\phi )$ as 
an integral operator, Nakanishi uses the following expression:
\begin{equation}
      \delta _{\rm N}(\phi )= \int_{\gamma} \rmd E \,  
         \phi (E^*)^* \delta _{\rm N}(E-z_{\rm R}) \, ,
          \label{Nakscon1}
\end{equation} 
where
\begin{equation}
       \delta _{\rm N}(E-z_{\rm R}) = \frac{1}{2\pi \rmi} \left(
              \frac{1}{E^{(-)}-z_{\rm R}} - \frac{1}{E^{(+)}-z_{\rm R}}
              \right) \, ,
\end{equation}
and where the contour $\gamma$ is such that the integral in 
Eq.~(\ref{Nakscon1}) decomposes into two terms. The end points of the 
integration paths are the same for the two terms, namely, $0$ and 
$+\infty$. The 
integration path for the first term, $\frac{1}{E^{(-)}-z_{\rm R}}$, passes 
below $z_{\rm R}$, whereas the integration path for the second term,
$\frac{1}{E^{(+)}-z_{\rm R}}$, passes above $z_{\rm R}$. Adding the two 
terms we obtain 
\begin{equation}
      \int_{\gamma}\rmd E \,  \phi (E^*)^* \delta _{\rm N}(E-z_{\rm R})=
      \frac{1}{2\pi \rmi} \oint \rmd E \, \frac{\phi (E^*)^*}{E-z_{\rm R}} =
        \phi (z_{\rm R}^*)^* \, .
          \label{Nakscon}
\end{equation} 
Thus, the distributional definition~(\ref{Cdelta-anti}) is equivalent 
to Nakanishi's definition~(\ref{Nakscon1})-(\ref{Nakscon}), because both 
approaches associate the same complex number, $\phi (z_{\rm R}^*)^*$, with 
the same test function, $\phi$.


\begin{thebibliography}{99}
\bibitem{NOTE1} The cross section will actually show a distinctive,
almost-Lorentzian peak if, in addition to $z_{\rm R}$ being a pole of the 
$S$ matrix, the resonance is not too broad ($\Gamma _{\rm R}/E_{\rm R} << 1$), 
and if other possible resonances are not in the vicinity of $z_{\rm R}$, so 
their interference with $z_{\rm R}$ can be ignored.
\end{thebibliography}
\end{document}